\documentclass[aps,prd,showpacs,preprintnumbers,twocolumn,groupedaddress,nofootinbib]{revtex4-1}

\usepackage[]{amssymb}
\usepackage[]{amsmath}
\usepackage[]{mathrsfs}
\usepackage[]{eucal}
\usepackage[]{tensor}
\usepackage[]{amsthm}
\usepackage[]{bm}

\newcommand{\dd}{\partial}
\newcommand{\df}{\mathrm{d}}

\newcommand{\Lie}{\pounds}
\newcommand{\FF}{\mathcal{F}}
\newcommand{\GG}{\mathcal{G}}
\newcommand{\LL}{\mathscr{L}}
\newcommand{\nab}[1]{\nabla_{\!#1}}
\newcommand{\qqd}{\ , \quad}

\newcommand{\bc}{\begin{center}}
\newcommand{\ec}{\end{center}}
\newcommand{\be}{\begin{equation}}
\newcommand{\ee}{\end{equation}}

\usepackage{dsfont}

\newcommand{\cl}[1]{\overline{#1}}

\usepackage[unicode]{hyperref}
\usepackage{xcolor}
\definecolor{pastgreen}{HTML}{669900}
\definecolor{pastblue}{HTML}{336699}
\definecolor{linkcol}{HTML}{663333}
\hypersetup{colorlinks,linkcolor={pastblue},citecolor={pastblue},urlcolor={pastblue}}

\theoremstyle{plain} \newtheorem{tm}{Theorem}[section]
\theoremstyle{plain} \newtheorem{lm}[tm]{Lemma}
\theoremstyle{plain} \newtheorem{cor}[tm]{Corollary}
\theoremstyle{definition} \newtheorem{defn}[tm]{Definition}
\theoremstyle{definition} 
\newcommand{\btm}{\begin{tm}}
\newcommand{\etm}{\end{tm}}
\newcommand{\blm}{\begin{lm}}
\newcommand{\elm}{\end{lm}}
\newcommand{\bcor}{\begin{cor}}
\newcommand{\ecor}{\end{cor}}
\newcommand{\bdefn}{\begin{defn}}
\newcommand{\edefn}{\end{defn}}

\begin{document}

\preprint{ZTF-EP-17-05}

\title{Nonlinear electromagnetic fields and symmetries}

\author{Irena Barja\v si\'c}
\email[]{ibarjasi@dominis.phy.hr}
\author{Luka Gulin}
\email[]{lgulin@dominis.phy.hr}
\author{Ivica Smoli\'c}
\email[]{ismolic@phy.hr}

\affiliation{Department of Physics, Faculty of Science, University of Zagreb, 10000 Zagreb, Croatia}

\date{\today}

\begin{abstract}
We extend the classical results on the symmetry inheritance of the canonical electromagnetic fields, described by the Maxwell's Lagrangian, to a much wider class of models, which include those of the Born-Infeld, power Maxwell and the Euler-Heisenberg type. Symmetry inheriting fields allow the introduction of electromagnetic scalar potentials and these are proven to be constant on the Killing horizons. Finally, using the relations obtained along the analysis, we generalize and simplify the recent proof for the symmetry inheritance of the 3-dimensional case, as well as give the first constraint for the higher dimensional electromagnetic fields.
\end{abstract}


\maketitle

%%%%%%%%%%%%%%%%%%%%%%%%%%%%%%%%%%%%%%%%%%%%%%%%%%%%%%%%%
%%%%%%%%%%%%%%%%%%%%%%%%%%%%%%%%%%%%%%%%%%%%%%%%%%%%%%%%%
\section{Introduction}%%%
%%%%%%%%%%%%%%%%%%%%%%%%%%%%%%%%%%%%%%%%%%%%%%%%%%%%%%%%%
%%%%%%%%%%%%%%%%%%%%%%%%%%%%%%%%%%%%%%%%%%%%%%%%%%%%%%%%%

Whenever we build a model of two classical relativistic interactions, gravitational and electro\-mag\-net\-ic, and analyse its solutions with symmetries, one of the basic initial questions is whether the gauge field shares the symmetries with the spacetime metric. The usual assumption in a typical choice of Ansatz, as well as in the various uniqueness theorems \cite{Heusler,HCC}, is that the electromagnetic field \emph{inherits} the spacetime symmetries. This assumption, however, deserves a justification as we know that there are solutions in which the symmetry inheritance is broken \cite{SKMHH}.

Let us assume that the spacetime $(M,g_{ab},F_{ab})$ is a smooth connected $D$-dimensional Lorentzian manifold, with the spacetime metric $g_{ab}$ and the electromagnetic tensor $F_{ab}$, which are solutions of the system of field equations derived from the Lagrangian of the form $L = L_{\mathrm{G}} + L_{\mathrm{EM}}$. Furthermore, in order to obtain the conclusions as general as possible, we shall assume that the gravitational field equation takes the form
\be\label{eq:EOM}
E_{ab} = 8\pi T_{ab} \ ,
\ee
where the tensor $E_{ab}$ is a general smooth function of the spacetime metric, the Riemann tensor, its covariant derivatives and the Levi-Civita tensor. In other words, the gravitational side of the equation (\ref{eq:EOM}), apart from the ordinary Einstein tensor with the cosmological constant, might be for example the Lovelock tensor \cite{Love71}, a tensor which is a member of the $f(R)$ theories \cite{DeFT10} or the generalized Cotton tensor, originating from the gravitational Chern-Simons terms \cite{BCPPS11a}. The electromagnetic energy-momentum tensor for the canonical, Maxwell's choice of Lagrangian 
\be\label{eq:LMax}
L_{\mathrm{EM}}^{\mathrm{(Max)}} = -\frac{1}{4}\,F_{ab} F^{ab}\,{*1}
\ee
is given by
\be\label{eq:TMax}
T_{ab}^{\mathrm{(Max)}} = \frac{1}{4\pi} \left( F_{ac} \tensor{F}{_b^c} - \frac{1}{4}\,g_{ab} F_{cd} F^{cd}\right) \ .
\ee
Now, if this spacetime allows at least one (sufficiently smooth) Killing vector field $\xi^a$, such that $\Lie_\xi g_{ab} = 0$, due to our assumptions we immediately have $\Lie_\xi E_{ab} = 0$ and the field equation (\ref{eq:EOM}) implies that
\be\label{eq:LieT}
\Lie_\xi T_{ab} = 0 \ .
\ee
From here we want to conclude the general form of the Lie derivative $\Lie_\xi F_{ab}$, that is, the symmetry inheritance properties of the electromagnetic field. Due to the series of papers \cite{Woo73a,*Woo73b,*MzHRS74,*RT75,*Coll75,MW75,WY76a,WY76b,Tod06} we know that the electromagnetic field in 4-dimensional spacetime can break the symmetry inheritance, but precisely such that
\be\label{eq:fourLieF}
\Lie_\xi F_{ab} = f\,{*F}_{ab} \ ,
\ee
for some real function $f$, which is a constant if $F_{ab}$ is non-null. There are several known examples \cite{LP77,MW75,WY76b,FC78} of the electrovac spacetimes with nonvanishing function $f$. Away from the dimension 4 the relation (\ref{eq:fourLieF}) cannot hold simply because the Hodge dual of the electromagnetic tensor $F_{ab}$ is a $(D-2)$-form, 
\be
{*F}_{a_1 \cdots a_{D-2}} = \frac{1}{2}\,F_{cd}\,\tensor{\epsilon}{^c^d_{a_1}_{\cdots}_{a_{D-2}}} \ .\ee
The generalization of the (\ref{eq:fourLieF}) to the higher dimensional, $D > 4$ cases is a long-standing open problem.

\medskip 

Almost a century ago, Born and Infeld \cite{Born34,*BI34} begun investigating alternative, nonlinear models of electromagnetic fields, which could cure the inconsistencies of the Maxwell's electrodynamics associated with the infinite self-energy of the point charges. Over the time it has been realized that these kind of models can be also used to regularize the black hole singularities \cite{ABG98,Bronnikov00,BH02} and the cosmological singularities \cite{GSB00,*GSB04,*CGMCL04}, as well as to simulate the ``dark energy'' \cite{ELNO03,*LR10}. Nonlinearities in the electromagnetic field appear in the quantum corrections to classical electromagnetic interaction \cite{HE36} and in the low energy effective Lagrangians of the superstring models \cite{FT85}. There are numerous experiments which will probe the nonlinearities in the electromagnetic fields \cite{BR13,FBR16,MCLP17,EMY17}.

\medskip 

Our objective is to find the constraints on the possible forms of breaking of the symmetry inheritance by the nonlinear electromagnetic fields and the conditions under which the symmetry inheritance is necessary. This is motivated by a broad range of questions, from the formal ones (removal of the unnecessary assumptions in general theorems \cite{Rasheed97,Bronnikov00,GHLM12,*CPX14}) to those aimed toward phenomenology (the possibility that the symmetry noninheriting fields might evade the no-hair theorems \cite{HR15,ISm17}). The nonlinear models of the electromagnetic field can be described by the general Lagrangian of the form
\be\label{eq:LagNLE}
L_{\mathrm{EM}} = \LL(\FF,\GG)\,{*1} \ ,
\ee
where we have introduced two standard electromagnetic invariants, 
\be
\FF \equiv F_{ab} F^{ab} \qquad \mathrm{and} \qquad \GG \equiv F_{ab}\,{*F}^{ab} \ .
\ee
Note that while the invariant $\FF$ is a scalar in any number of dimensions, the invariant $\GG$ is a scalar only in 4-dimensional spacetimes! For the derivatives we shall use abbreviations, such as
$$\LL_\FF \equiv \frac{\dd\LL}{\dd\FF} \qqd \LL_\GG \equiv \frac{\dd\LL}{\dd\GG} \qqd \LL_{\FF\FF} \equiv \frac{\dd^2\LL}{\dd\FF^2} \qqd \mathrm{etc.}$$
Most notable examples of such Lagrangians are

\begin{itemize}

\item[(i)] the Born-Infeld \cite{Born34,BI34,GSP84},
\be\label{eq:BI}
\LL^{\mathrm{(BI)}}(\FF,\GG) = b^2 \left( 1 - \sqrt{1 + \frac{\FF}{2b^2} - \frac{\GG^2}{16b^4}} \right)
\ee
for some real constant $b > 0$;

\item[(ii)] the ``power Maxwell'' \cite{HM07,*HM08},
\be
\LL^{\mathrm{(pM)}}(\FF) = C\,\FF^s
\ee
with some real constants $C > 0$ and $s \ne 0$;

\item[(iii)] the Euler-Heisenberg \cite{HE36,YT00,RWX13},
\be\label{eq:EH}
\LL^{\mathrm{(EH)}}(\FF,\GG) = -\frac{1}{4}\,\FF + \gamma (4\FF^2 + 7\GG^2) + \, \dots
\ee
with some real constant $\gamma > 0$.
\end{itemize}

Along with these the literature abounds with various other nonlinear models, such as the Hoffmann-Infeld \cite{HI37,AFG05}, the logarithmic \cite{Soleng95,Hendi12}, the exponential \cite{Hendi12}, those based on a rational function \cite{Kruglov15}, etc. 

\medskip

The energy-momentum tensor corresponding to (\ref{eq:LagNLE}) is given by
\be\label{eq:NLT}
T_{ab} = -\frac{1}{4\pi} \Big( (\LL_\GG\,\GG - \LL) g_{ab} + 4 \LL_\FF \, F_{ac} \tensor{F}{_b^c} \Big) \ .
\ee
The generalized Maxwell's equations are
\be\label{eq:gNLMax}
\df F = 0 \qquad \mathrm{and} \qquad \df\,{*Z} = 0 \ ,
\ee
where we have introduced an auxiliary two-form
\be\label{eq:Z}
Z_{ab} \equiv -4 \big( \LL_\FF\,F_{ab} + \LL_\GG\,{*F}_{ab} \big) \ .
\ee
The normalization is chosen so that $Z_{ab}$ reduces to $F_{ab}$ in the Maxwell's case.

%%%%%%%%%%%%%%%%%%%%%%%%%%%%%%%%%%%%%%%%%%%%%%%%%%%%%%%%%
%%%%%%%%%%%%%%%%%%%%%%%%%%%%%%%%%%%%%%%%%%%%%%%%%%%%%%%%%
\section{Symmetry inheritance}%%%
%%%%%%%%%%%%%%%%%%%%%%%%%%%%%%%%%%%%%%%%%%%%%%%%%%%%%%%%%
%%%%%%%%%%%%%%%%%%%%%%%%%%%%%%%%%%%%%%%%%%%%%%%%%%%%%%%%%

Throughout the paper we assume that all the fields are sufficiently smooth, and focus on the open set of points $O \subseteq M$ where $\LL_\FF \ne 0$. Let us first assume that $D = 4$. Using the trace $T \equiv g^{ab} T_{ab}$ we may write the energy-tensor in a convenient way (see e.g.~\cite{GH01}),
\be\label{eq:TTMax}
T_{ab} = - 4\LL_\FF\,T_{ab}^{\mathrm{(Max)}} + \frac{1}{4}\,T g_{ab} \ .
\ee
From here, with the master equation (\ref{eq:LieT}), we get
\be\label{eq:LieLFT}
\Lie_\xi (\LL_\FF\,T_{ab}^{\mathrm{(Max)}}) = 0 \ .
\ee
On the set $O$ this equation is nontrivial and we may introduce an auxiliary field 
\be
\widetilde{F}_{ab} \equiv \sqrt{|\LL_\FF|} F_{ab} \ ,
\ee
so that (\ref{eq:LieLFT}) becomes
\be
\Lie_\xi \left( \widetilde{F}_{ac} \tensor{\widetilde{F}}{_b^c} - \frac{1}{4}\,g_{ab}\,\widetilde{F}_{pq} \widetilde{F}^{pq} \right) = 0 \ .
\ee
In this way we have formally reduced the problem to the symmetry inheritance in the Maxwell's case (note that in the set $O$ the tensor $F_{ab}$ is (non-)null if and only if $\widetilde{F}_{ab}$ is (non-)null). Whence, using the classical results \cite{MW75,RT75,Coll75,WY76a,WY76b,Tod06} (see also section 11.1 in \cite{SKMHH}), we can conclude that 
\be
\Lie_\xi \widetilde{F}_{ab} = \alpha\,{*\widetilde{F}}_{ab}
\ee
for some real function $\alpha$ and, expressing everything with the original field $F_{ab}$, we have the following theorem.

\btm
Let $(M,g_{ab},F_{ab})$ be a $(1+3)$-di\-men\-sional solution of (\ref{eq:EOM}) and (\ref{eq:gNLMax}), allowing a (sufficiently smooth) Killing vector field $\xi^a$. Then on the set of points where $\LL_\FF \ne 0$ we have
\be\label{eq:LieF}
\Lie_\xi F_{ab} = \alpha\,{*F}_{ab} + \beta\,F_{ab} \ ,
\ee
where $\alpha$ and $\beta$ are real functions, such that
\be
\beta = -\frac{1}{2\LL_\FF}\,\Lie_\xi \LL_\FF \ .
\ee
If $\alpha = 0$ and $\FF \LL_{\FF\FF} + \LL_\FF \ne 0$ then $\beta = 0$.
\etm

Suppose now that the spacetime admits an $r$-parameter group of isometries, generated by the Killing vector fields $\xi^a_{(i)}$ which satisfy the commutation relations
\be
[\xi_{(i)},\xi_{(j)}]^a = C^k_{ij}\,\xi_{(k)}^a
\ee
with the group structure constants $C^k_{ij}$. For each of these Killing vector fields we have  
\be
\Lie_{\xi_{(i)}} F_{ab} = \alpha_{(i)}\,{*F}_{ab} + \beta_{(i)}\,F_{ab} \ .
\ee
Using the elementary property of the Lie derivatives,
\be
\Lie_{[X,Y]} = [\Lie_X,\Lie_Y]
\ee
we get the equation
\be
C^k_{ij} \Lie_{\xi_{(k)}} F_{ab} = C^k_{ij} \beta_{(k)} F_{ab} + ( \Lie_{\xi_{(i)}} \alpha_{(j)} - \Lie_{\xi_{(j)}} \alpha_{(i)} ) {*F}_{ab} \ ,
\ee
from where it follows that at each point where $F_{ab} \ne 0$ we have a linear system of relations 
\be\label{eq:Calpha}
C^k_{ij} \alpha_{(k)} = \Lie_{\xi_{(i)}} \alpha_{(j)} - \Lie_{\xi_{(j)}} \alpha_{(i)} \ .
\ee
The analysis of these constraints on the functions $\alpha_{(k)}$ is left for the future work.

\medskip

In the rest of the section we shall look more closely at the special case when $\LL = \LL(\FF)$. In order to put it in the broader perspective we shall first assume that the number of spacetime dimensions is some general $D \ge 2$. The energy-momentum tensor now takes the form
\be
T_{ab} = \frac{1}{4\pi} \Big( \LL\,g_{ab} - 4\LL_\FF\,F_{ac}\tensor{F}{_b^c} \Big) \ .
\ee
Using $\Lie_\xi T = 0$ we have
\be\label{eq:KLieF}
0 = \Big( 4\FF\,\LL_{\FF\FF} - (D-4)\LL_\FF \Big) \Lie_\xi \FF \equiv \mathcal{K}(\FF) \Lie_\xi \FF \ .
\ee
Let us denote by $V \subseteq O$ the set of points in which the function $\mathcal{K}(\FF)$ does not vanish or which are elements of the open sets on which $\FF$ is constant and equal to a zero of the function $\mathcal{K}(\FF)$. We shall refer to the elements of the set $W = \cl{V} \cap O$ as the \emph{regular points} of $O$. At each point of $W$ we immediately have $\Lie_\xi \FF = 0$ and, consequently, 
\be\label{eq:LieLLieLF}
\Lie_\xi \LL = \LL_\FF \Lie_\xi \FF = 0 \quad \mathrm{and} \quad \Lie_\xi \LL_\FF = \LL_{\FF\FF} \Lie_\xi \FF = 0 \ .
\ee
If $\mathcal{K}(\FF)$ vanishes for \emph{any} $\FF$ then we cannot extract any useful information from Eq.~(\ref{eq:KLieF}). Equation $\mathcal{K}(\FF) = 0$ is an ordinary differential equation with the general solution of the form $\LL(\FF) = A \FF^{D/4} + B$ for some real constants $A$ and $B$. The constant $B$ only contributes to the cosmological constant, so we can dismiss it in this discussion. Therefore, the ``blind spot'' of the analysis are the Lagrangians of the form $L_{\mathrm{EM}} = A\FF^{D/4}\,{*1}$, those for which the electromagnetic field action becomes conformally invariant and the corresponding energy-momentum traceless (the choice which was exploited in \cite{HM07}). All the remaining nonlinear Lagrangians mentioned in this paper are devoid of irregular points of the set $O$.

One of the consequences of the relations (\ref{eq:LieLLieLF}) is that on the set $W$ the equation (\ref{eq:LieT}) implies that in fact
\be\label{eq:LieFFab}
\Lie_\xi (F_{ac} \tensor{F}{_b^c}) = 0 \ .
\ee
To our knowledge this is the first concrete constraint on the symmetry inheritance properties of the higher dimensional electromagnetic fields. Unfortunately, we have been unable to find any other relation which would provide substantial information on the nature of $\Lie_\xi F_{ab}$ in $D > 4$.

\medskip

In the $D=4$ case the relations (\ref{eq:LieLLieLF}) imply that $\beta = 0$ in (\ref{eq:LieF}), and $0 = \Lie_\xi \FF = 2\alpha\GG$ holds at each point of the set $W$. Thus, either $\alpha = 0$ (in which case the symmetry is inherited) or $\GG = 0$. Furthermore, using $\Lie_\xi \GG = -2\alpha\FF$, we see that on the interior of points where $\alpha \ne 0$ and $\GG = 0$ we necessarily have $\FF = 0$. In other words,  either the electromagnetic field is null or it must inherit the spacetime symmetries. We can exclude such symmetry noninheriting null electromagnetic fields at least in a static spacetime. First, at each point of the set $O$ we have (see Eq.~(\ref{eq:TTMax})),
\be
\xi_{[a}\,T_{b]c}\,\xi^c = 0 \quad \mathrm{iff} \quad \xi_{[a}\,T_{b]c}^{\mathrm{(Max)}}\,\xi^c = 0 \ .
\ee 
So, using the well-known theorems \cite{Banerjee70,Tod06}, if the tensor $E_{ab}$ in (\ref{eq:EOM}) belongs to the orthogonal-transitive class of order 1 \cite{ISm17}, then in every static subset of $O$, namely all the points where $\xi^a$ is timelike and satisfies the Frobenius condition $\xi_{[a} \nab{b}\,\xi_{c]} = 0$, the electromagnetic tensor is either trivial, $F_{ab} = 0$, or non-null.

\medskip

We already have examples (\cite{LP77}; ``Example 1'' in \cite{WY76b}; \cite{FC78}) of exact solutions of the Einstein-Maxwell field equations with the symmetry noninheriting null electromagnetic fields. A simple way to ``recycle'' these solutions in the nonlinear case is to look at those models for which the energy-momentum tensor (\ref{eq:NLT}) reduces to the Maxwell's energy-momentum tensor (\ref{eq:TMax}) and the electromagnetic tensor $F_{ab}$ reduces to the tensor $Z_{ab}$ for the null electromagnetic fields. Such models can be found among those whose Lagrangian density satisfies
\be\label{eq:Maxasympt}
\lim_{\FF \to 0} \LL(\FF) = 0 \quad \mathrm{and} \quad \lim_{\FF \to 0} \LL_\FF(\FF) = -\frac{1}{4} \ .
\ee
These conditions, usually referred to as Maxwell's asymptotics \cite{Bronnikov00}, at the same time guarantee the physical admissibility of the nonlinear models. For example, the truncated version of the Born-Infeld model (in which the $b^{-4}$ term in (\ref{eq:BI}) is suppressed) satisfies the conditions (\ref{eq:Maxasympt}), thus all the Einstein-Maxwell symmetry noninheriting null electromagnetic fields are automatically solutions of the Einstein-Born-Infeld field equations.

%%%%%%%%%%%%%%%%%%%%%%%%%%%%%%%%%%%%%%%%%%%%%%%%%%%%%%%%%
%%%%%%%%%%%%%%%%%%%%%%%%%%%%%%%%%%%%%%%%%%%%%%%%%%%%%%%%%
\section{The lower dimensional cases}%%%
%%%%%%%%%%%%%%%%%%%%%%%%%%%%%%%%%%%%%%%%%%%%%%%%%%%%%%%%%
%%%%%%%%%%%%%%%%%%%%%%%%%%%%%%%%%%%%%%%%%%%%%%%%%%%%%%%%%

Lower dimensional spacetimes are often used as a toy models, whence for the completeness we turn our attention to them, assuming that $\LL = \LL(\FF)$.

\medskip

\emph{$(1+1)$-dimensional case}. As $F_{ab}$ is a form of maximal rank here, we have $F_{ab} = f \epsilon_{ab}$ for some function $f$ and thus $\Lie_\xi F_{ab} = (\Lie_\xi f) \epsilon_{ab}$. One of the Maxwell's equations, $\df F = 0$, is automatically satisfied, while the other implies that $\Lie_\xi(\LL_\FF f) = 0$. In the canonical case (\ref{eq:LMax}) we immediately have $\Lie_\xi f = 0$, while in the nonlinear case the same conclusion holds on the set $W$, where we have both $\Lie_\xi \LL_\FF = 0$ and $\LL_\FF \ne 0$. Hence, in both cases we have the symmetry inheritance.

\medskip

\emph{$(1+2)$-dimensional case}. Recently it has been shown \cite{CDPS16}, using decomposition of $F_{ab}$ to the electric and the magnetic parts, that the 3-dimensional Maxwell's electromagnetic field (with possible presence of the gauge Chern-Simons terms) necessarily inherits the spacetime symmetries. In order to attack the problem for the nonlinear Lagrangians which allow a nonempty set $W$ of regular points we resort to different strategy. Along a neighbourhood of an orbit of the Killing vector field $\xi^a$ one can introduce a dreibein basis,
\be
g_{ab} e^a_{(\mu)} e^b_{(\nu)} = \eta_{\mu\nu} = \mathrm{diag}(-1,+1,+1) \ ,
\ee
Lie-dragged along the field $\xi^a$, such that $\Lie_\xi e^a_{(\mu)} = 0$. Then we have a decomposition of the electromagnetic tensor,
\be\label{eq:Fgamgam}
F_{ab} = \gamma_{ij} e_a^{(i)} e_b^{(j)} \ ,
\ee
where $\gamma_{(ij)} = 0$, and a decomposition of the relation (\ref{eq:LieFFab}),
\be\label{eq:FFsigma}
F_{ac} \tensor{F}{_b^c} = \sigma_{ij} e_a^{(i)} e_b^{(j)} \ ,
\ee
where $\sigma_{[ij]} = 0$ and $\Lie_\xi \sigma_{ij} = 0$ for each $i$ and $j$ ($\gamma_{ij}$ and $\sigma_{ij}$ should be taken just as a set of auxiliary functions). By combining (\ref{eq:Fgamgam}) and (\ref{eq:FFsigma}), we get a system of equations,
\be
\eta^{kl}\gamma_{ik}\gamma_{jl} = \sigma_{ij} \ ,
\ee
from which we can express the functions $\gamma_{ij}$ with the functions $\sigma_{ij}$. This implies that $\Lie_\xi \gamma_{ij} = 0$ for each $i$ and $j$, thus $\Lie_\xi F_{ab} = 0$ at least on the set $W$.

Yet another proof of the symmetry inheritance in the 3-dimensional case can be obtained via correspondence with the real scalar field \cite{KT17}, in combination with the recent general results on symmetry inheritance of the scalar fields \cite{ISm15,ISm17}.

%%%%%%%%%%%%%%%%%%%%%%%%%%%%%%%%%%%%%%%%%%%%%%%%%%%%%%%%%
%%%%%%%%%%%%%%%%%%%%%%%%%%%%%%%%%%%%%%%%%%%%%%%%%%%%%%%%%
\section{Electromagnetic scalar potentials}%%%
%%%%%%%%%%%%%%%%%%%%%%%%%%%%%%%%%%%%%%%%%%%%%%%%%%%%%%%%%
%%%%%%%%%%%%%%%%%%%%%%%%%%%%%%%%%%%%%%%%%%%%%%%%%%%%%%%%%

Given a Killing vector field $\xi^a$ one can define the electric 1-form $E = -i_\xi F$ and the magnetic 1-form $H = i_\xi {*Z}$. Furthermore, whenever this is possible, it is convenient to introduce the electric scalar potential $\Phi$ and the magnetic scalar potential $\Psi$, such that $E = \df\Phi$ and $H = \df\Psi$. Now, since $\df E = -\Lie_\xi F$, the symmetry inheritance is \emph{necessary} for the existence of $\Phi$ (and sufficient at least to guarantee its local existence). Also, as $\df H = *\Lie_\xi Z$, the symmetry inheritance is sufficient to guarantee the local existence of the potential $\Psi$.

One of the basic building blocks in the first law of black hole mechanics and in the various black hole uniqueness theorems \cite{Heusler} is the proof of the constancy of $\Phi$ and $\Psi$ on the black hole horizons. There are various strategies (see \cite{ISm12,*ISm14} for an overview) to prove this zeroth law of the black hole electrodynamics, each with different benefits and disadvantages: by assuming that $E_{ab}$ is the Einstein's tensor (this was exploited in \cite{Rasheed97}), by assuming that the black hole horizon is of the bifurcate type (which immediately works for the nonlinear electrodynamics) and the approach based on the symmetry \cite{ISm12,ISm14} (which is independent of the gravitational field equations or the presence of the bifurcation surface). Let us examine more closely the third approach in the context of the nonlinear electrodynamics.

For example, assume that the spacetime is circular: stationary, axisymmetric, with the corresponding commuting Killing vector fields $k^a$ and $m^a$ (the latter is the axial Killing vector with compact orbits), which satisfy the Frobenius conditions,
\be
k_{[a} m_b \nab{c} k_{d]} = k_{[a} m_b \nab{c} m_{d]} = 0 \ .
\ee
If we use the identity
\be
i_X \Lie_Y - i_Y \Lie_X = i_X i_Y \df - \df i_X i_Y + i_{[X,Y]}
\ee
with the Killing vector fields, $X^a = k^a$ and $Y^a = m^a$, and apply it on $F_{ab}$ and ${*Z}_{ab}$, we get that both $F(k,m)$ and $*Z(k,m)$ are constant. Thus, on any connected domain of the spacetime which contains the points where any of these two Killing vector fields vanish (such is the axis where $m^a = 0$) we have 
\be
F(k,m) = 0 = {*Z}(k,m) \ .
\ee 
This allows us to apply the method presented in \cite{ISm12,ISm14} to prove that both $\Phi$ and $\Psi$ are constant on any connected component of the Killing horizon $H[\chi]$, generated by the Killing vector field $\chi^a = k^a + \Omega_{\mathsf{H}} m^a$ (where $\Omega_{\mathsf{H}}$ plays the role of the ``angular velocity of the horizon'' \cite{Heusler}).

%%%%%%%%%%%%%%%%%%%%%%%%%%%%%%%%%%%%%%%%%%%%%%%%%%%%%%%%
%%%%%%%%%%%%%%%%%%%%%%%%%%%%%%%%%%%%%%%%%%%%%%%%%%%%%%%%
\section{Final remarks}%%%
%%%%%%%%%%%%%%%%%%%%%%%%%%%%%%%%%%%%%%%%%%%%%%%%%%%%%%%%
%%%%%%%%%%%%%%%%%%%%%%%%%%%%%%%%%%%%%%%%%%%%%%%%%%%%%%%%

Our results should be taken as a guiding blueprint for the symmetry noninheriting nonlinear electromagnetic fields, examples of which are still few and far between. For example, it is an open question if the models built upon the Born-Infeld (\ref{eq:BI}) or the Euler-Heisenberg Lagrangian (\ref{eq:EH}) allow such fields in the cases when the field equations do not reduce simply to the Einstein-Maxwell's. The higher dimensional fields are just weakly constrained with respect to the symmetry inheritance and here one could expect new surprises. Finally, it would be interesting to look for the further constraints imposed by some specific boundary conditions, either from the presence of the black hole horizons or the asymptotic conditions at infinity \cite{Tod06}.

\begin{acknowledgments}
This research has been supported by the Croatian Science Foundation under the project No.~8946.
\end{acknowledgments}

%%%%%%%%%%%%%%%%%%%%%%%%%%%%
%%%%%%%%%%%%%%%%%%%%%%%%%%%%
\bibliography{nlemsym}

\begin{thebibliography}{55}%
\makeatletter
\providecommand \@ifxundefined [1]{%
 \@ifx{#1\undefined}
}%
\providecommand \@ifnum [1]{%
 \ifnum #1\expandafter \@firstoftwo
 \else \expandafter \@secondoftwo
 \fi
}%
\providecommand \@ifx [1]{%
 \ifx #1\expandafter \@firstoftwo
 \else \expandafter \@secondoftwo
 \fi
}%
\providecommand \natexlab [1]{#1}%
\providecommand \enquote  [1]{``#1''}%
\providecommand \bibnamefont  [1]{#1}%
\providecommand \bibfnamefont [1]{#1}%
\providecommand \citenamefont [1]{#1}%
\providecommand \href@noop [0]{\@secondoftwo}%
\providecommand \href [0]{\begingroup \@sanitize@url \@href}%
\providecommand \@href[1]{\@@startlink{#1}\@@href}%
\providecommand \@@href[1]{\endgroup#1\@@endlink}%
\providecommand \@sanitize@url [0]{\catcode `\\12\catcode `\$12\catcode
  `\&12\catcode `\#12\catcode `\^12\catcode `\_12\catcode `\%12\relax}%
\providecommand \@@startlink[1]{}%
\providecommand \@@endlink[0]{}%
\providecommand \url  [0]{\begingroup\@sanitize@url \@url }%
\providecommand \@url [1]{\endgroup\@href {#1}{\urlprefix }}%
\providecommand \urlprefix  [0]{URL }%
\providecommand \Eprint [0]{\href }%
\providecommand \doibase [0]{http://dx.doi.org/}%
\providecommand \selectlanguage [0]{\@gobble}%
\providecommand \bibinfo  [0]{\@secondoftwo}%
\providecommand \bibfield  [0]{\@secondoftwo}%
\providecommand \translation [1]{[#1]}%
\providecommand \BibitemOpen [0]{}%
\providecommand \bibitemStop [0]{}%
\providecommand \bibitemNoStop [0]{.\EOS\space}%
\providecommand \EOS [0]{\spacefactor3000\relax}%
\providecommand \BibitemShut  [1]{\csname bibitem#1\endcsname}%
\let\auto@bib@innerbib\@empty
%</preamble>
\bibitem [{\citenamefont {Heusler}(1996)}]{Heusler}%
  \BibitemOpen
  \bibfield  {author} {\bibinfo {author} {\bibfnamefont {M.}~\bibnamefont
  {Heusler}},\ }\href@noop {} {\emph {\bibinfo {title} {{Black Hole Uniqueness
  Theorems}}}}\ (\bibinfo  {publisher} {Cambridge University Press},\ \bibinfo
  {address} {Cambridge New York},\ \bibinfo {year} {1996})\BibitemShut
  {NoStop}%
\bibitem [{\citenamefont {Chru{\'s}ciel}\ \emph {et~al.}(2012)\citenamefont
  {Chru{\'s}ciel}, \citenamefont {Lopes~Costa},\ and\ \citenamefont
  {Heusler}}]{HCC}%
  \BibitemOpen
  \bibfield  {author} {\bibinfo {author} {\bibfnamefont {P.~T.}\ \bibnamefont
  {Chru{\'s}ciel}}, \bibinfo {author} {\bibfnamefont {J.}~\bibnamefont
  {Lopes~Costa}}, \ and\ \bibinfo {author} {\bibfnamefont {M.}~\bibnamefont
  {Heusler}},\ }\href {\doibase 10.12942/lrr-2012-7} {\bibfield  {journal}
  {\bibinfo  {journal} {Living Rev. Rel.}\ }\textbf {\bibinfo {volume} {{\bf
  15}}},\ \bibinfo {pages} {7} (\bibinfo {year} {2012})},\ \Eprint
  {http://arxiv.org/abs/1205.6112} {arXiv:1205.6112 [gr-qc]} \BibitemShut
  {NoStop}%
%%CITATION = ARXIV:1205.6112;%%
\bibitem [{\citenamefont {Stephani}\ \emph {et~al.}(2003)\citenamefont
  {Stephani}, \citenamefont {Kramer}, \citenamefont {MacCallum}, \citenamefont
  {Hoenselaers},\ and\ \citenamefont {Herlt}}]{SKMHH}%
  \BibitemOpen
  \bibfield  {author} {\bibinfo {author} {\bibfnamefont {H.}~\bibnamefont
  {Stephani}}, \bibinfo {author} {\bibfnamefont {D.}~\bibnamefont {Kramer}},
  \bibinfo {author} {\bibfnamefont {M.}~\bibnamefont {MacCallum}}, \bibinfo
  {author} {\bibfnamefont {C.}~\bibnamefont {Hoenselaers}}, \ and\ \bibinfo
  {author} {\bibfnamefont {E.}~\bibnamefont {Herlt}},\ }\href@noop {} {\emph
  {\bibinfo {title} {Exact Solutions of Einstein's Field Equations}}}\
  (\bibinfo  {publisher} {Cambridge University Press},\ \bibinfo {address}
  {Cambridge, England},\ \bibinfo {year} {2003})\BibitemShut {NoStop}%
\bibitem [{\citenamefont {Lovelock}(1971)}]{Love71}%
  \BibitemOpen
  \bibfield  {author} {\bibinfo {author} {\bibfnamefont {D.}~\bibnamefont
  {Lovelock}},\ }\href {\doibase 10.1063/1.1665613} {\bibfield  {journal}
  {\bibinfo  {journal} {J. Math. Phys.}\ }\textbf {\bibinfo {volume} {{\bf
  12}}},\ \bibinfo {pages} {498} (\bibinfo {year} {1971})}\BibitemShut
  {NoStop}%
%%CITATION = JMAPA,12,498;%%
\bibitem [{\citenamefont {De~Felice}\ and\ \citenamefont
  {Tsujikawa}(2010)}]{DeFT10}%
  \BibitemOpen
  \bibfield  {author} {\bibinfo {author} {\bibfnamefont {A.}~\bibnamefont
  {De~Felice}}\ and\ \bibinfo {author} {\bibfnamefont {S.}~\bibnamefont
  {Tsujikawa}},\ }\href {\doibase 10.12942/lrr-2010-3} {\bibfield  {journal}
  {\bibinfo  {journal} {Living Rev. Rel.}\ }\textbf {\bibinfo {volume} {{\bf
  13}}},\ \bibinfo {pages} {3} (\bibinfo {year} {2010})},\ \Eprint
  {http://arxiv.org/abs/1002.4928} {arXiv:1002.4928 [gr-qc]} \BibitemShut
  {NoStop}%
%%CITATION = ARXIV:1002.4928;%%
\bibitem [{\citenamefont {Bonora}\ \emph {et~al.}(2011)\citenamefont {Bonora},
  \citenamefont {Cvitan}, \citenamefont {Dominis~Prester}, \citenamefont
  {Pallua},\ and\ \citenamefont {Smoli{\'c}}}]{BCPPS11a}%
  \BibitemOpen
  \bibfield  {author} {\bibinfo {author} {\bibfnamefont {L.}~\bibnamefont
  {Bonora}}, \bibinfo {author} {\bibfnamefont {M.}~\bibnamefont {Cvitan}},
  \bibinfo {author} {\bibfnamefont {P.}~\bibnamefont {Dominis~Prester}},
  \bibinfo {author} {\bibfnamefont {S.}~\bibnamefont {Pallua}}, \ and\ \bibinfo
  {author} {\bibfnamefont {I.}~\bibnamefont {Smoli{\'c}}},\ }\href {\doibase
  10.1007/JHEP07(2011)085} {\bibfield  {journal} {\bibinfo  {journal} {JHEP}\
  }\textbf {\bibinfo {volume} {{\bf 1107}}},\ \bibinfo {pages} {085} (\bibinfo
  {year} {2011})},\ \Eprint {http://arxiv.org/abs/1104.2523} {arXiv:1104.2523
  [hep-th]} \BibitemShut {NoStop}%
%%CITATION = ARXIV:1104.2523;%%
\bibitem [{\citenamefont {Woolley}(1973{\natexlab{a}})}]{Woo73a}%
  \BibitemOpen
  \bibfield  {author} {\bibinfo {author} {\bibfnamefont {M.}~\bibnamefont
  {Woolley}},\ }\href {http://projecteuclid.org/euclid.cmp/1103858928}
  {\bibfield  {journal} {\bibinfo  {journal} {Comm. Math. Phys.}\ }\textbf
  {\bibinfo {volume} {{\bf 31}}},\ \bibinfo {pages} {75} (\bibinfo {year}
  {1973}{\natexlab{a}})}\BibitemShut {NoStop}%
\bibitem [{\citenamefont {Woolley}(1973{\natexlab{b}})}]{Woo73b}%
  \BibitemOpen
  \bibfield  {author} {\bibinfo {author} {\bibfnamefont {M.}~\bibnamefont
  {Woolley}},\ }\href {http://projecteuclid.org/euclid.cmp/1103859250}
  {\bibfield  {journal} {\bibinfo  {journal} {Comm. Math. Phys.}\ }\textbf
  {\bibinfo {volume} {{\bf 33}}},\ \bibinfo {pages} {135} (\bibinfo {year}
  {1973}{\natexlab{b}})}\BibitemShut {NoStop}%
\bibitem [{\citenamefont {M{\"u}ller~zum Hagen}\ \emph
  {et~al.}(1974)\citenamefont {M{\"u}ller~zum Hagen}, \citenamefont
  {Robinson},\ and\ \citenamefont {Seifert}}]{MzHRS74}%
  \BibitemOpen
  \bibfield  {author} {\bibinfo {author} {\bibfnamefont {H.}~\bibnamefont
  {M{\"u}ller~zum Hagen}}, \bibinfo {author} {\bibfnamefont {D.}~\bibnamefont
  {Robinson}}, \ and\ \bibinfo {author} {\bibfnamefont {H.}~\bibnamefont
  {Seifert}},\ }\href {\doibase 10.1007/BF00758075} {\bibfield  {journal}
  {\bibinfo  {journal} {Gen. Relativ. Gravit.}\ }\textbf {\bibinfo {volume}
  {{\bf 5}}},\ \bibinfo {pages} {61} (\bibinfo {year} {1974})}\BibitemShut
  {NoStop}%
\bibitem [{\citenamefont {Ray}\ and\ \citenamefont {Thompson}(1975)}]{RT75}%
  \BibitemOpen
  \bibfield  {author} {\bibinfo {author} {\bibfnamefont {J.~R.}\ \bibnamefont
  {Ray}}\ and\ \bibinfo {author} {\bibfnamefont {E.~L.}\ \bibnamefont
  {Thompson}},\ }\href {\doibase doi:10.1063/1.522548} {\bibfield  {journal}
  {\bibinfo  {journal} {J. Math. Phys.}\ }\textbf {\bibinfo {volume} {{\bf
  16}}},\ \bibinfo {pages} {345} (\bibinfo {year} {1975})}\BibitemShut
  {NoStop}%
\bibitem [{\citenamefont {Coll}(1975)}]{Coll75}%
  \BibitemOpen
  \bibfield  {author} {\bibinfo {author} {\bibfnamefont {B.}~\bibnamefont
  {Coll}},\ }\href
  {http://gallica.bnf.fr/ark:/12148/bpt6k62167964/f357.item.r=1773} {\bibfield
  {journal} {\bibinfo  {journal} {C. R. Acad. Sci. (Paris) A}\ }\textbf
  {\bibinfo {volume} {{\bf 280}}},\ \bibinfo {pages} {1773} (\bibinfo {year}
  {1975})}\BibitemShut {NoStop}%
\bibitem [{\citenamefont {Michalski}\ and\ \citenamefont
  {Wainwright}(1975)}]{MW75}%
  \BibitemOpen
  \bibfield  {author} {\bibinfo {author} {\bibfnamefont {H.}~\bibnamefont
  {Michalski}}\ and\ \bibinfo {author} {\bibfnamefont {J.}~\bibnamefont
  {Wainwright}},\ }\href {\doibase 10.1007/BF00751574} {\bibfield  {journal}
  {\bibinfo  {journal} {Gen. Relativ. Gravit.}\ }\textbf {\bibinfo {volume}
  {{\bf 6}}},\ \bibinfo {pages} {289} (\bibinfo {year} {1975})}\BibitemShut
  {NoStop}%
\bibitem [{\citenamefont {Wainwright}\ and\ \citenamefont
  {Yaremovicz}(1976{\natexlab{a}})}]{WY76a}%
  \BibitemOpen
  \bibfield  {author} {\bibinfo {author} {\bibfnamefont {J.}~\bibnamefont
  {Wainwright}}\ and\ \bibinfo {author} {\bibfnamefont {P.~E.~A.}\ \bibnamefont
  {Yaremovicz}},\ }\href {\doibase 10.1007/BF00771105} {\bibfield  {journal}
  {\bibinfo  {journal} {Gen. Relativ. Gravit.}\ }\textbf {\bibinfo {volume}
  {{\bf 7}}},\ \bibinfo {pages} {345} (\bibinfo {year}
  {1976}{\natexlab{a}})}\BibitemShut {NoStop}%
%%CITATION = GRGVA,7,345;%%
\bibitem [{\citenamefont {Wainwright}\ and\ \citenamefont
  {Yaremovicz}(1976{\natexlab{b}})}]{WY76b}%
  \BibitemOpen
  \bibfield  {author} {\bibinfo {author} {\bibfnamefont {J.}~\bibnamefont
  {Wainwright}}\ and\ \bibinfo {author} {\bibfnamefont {P.~A.~E.}\ \bibnamefont
  {Yaremovicz}},\ }\href {\doibase 10.1007/BF00763408} {\bibfield  {journal}
  {\bibinfo  {journal} {Gen. Relativ. Gravit.}\ }\textbf {\bibinfo {volume}
  {{\bf 7}}},\ \bibinfo {pages} {595} (\bibinfo {year}
  {1976}{\natexlab{b}})}\BibitemShut {NoStop}%
%%CITATION = GRGVA,7,595;%%
\bibitem [{\citenamefont {Tod}(2007)}]{Tod06}%
  \BibitemOpen
  \bibfield  {author} {\bibinfo {author} {\bibfnamefont {P.}~\bibnamefont
  {Tod}},\ }\href {\doibase 10.1007/s10714-006-0363-5} {\bibfield  {journal}
  {\bibinfo  {journal} {Gen. Relativ. Gravit.}\ }\textbf {\bibinfo {volume}
  {{\bf 39}}},\ \bibinfo {pages} {111} (\bibinfo {year} {2007})},\ \Eprint
  {http://arxiv.org/abs/gr-qc/0611035} {arXiv:gr-qc/0611035 [gr-qc]}
  \BibitemShut {NoStop}%
%%CITATION = GR-QC/0611035;%%
\bibitem [{\citenamefont {Luk{\'a}cs}\ and\ \citenamefont
  {Perj{\'e}s}(1977)}]{LP77}%
  \BibitemOpen
  \bibfield  {author} {\bibinfo {author} {\bibfnamefont {B.}~\bibnamefont
  {Luk{\'a}cs}}\ and\ \bibinfo {author} {\bibfnamefont {Z.}~\bibnamefont
  {Perj{\'e}s}},\ }\enquote {\bibinfo {title} {{Time-Dependent Maxwell Fields
  in Stationary Geometry}},}\ in\ \href@noop {} {\emph {\bibinfo {booktitle}
  {Proceedings of the First Marcel Grossmann Meeting on General Relativity}}}\
  (\bibinfo  {publisher} {North-Holland Publishing Company, Amsterdam, New
  York, Oxford},\ \bibinfo {year} {1977})\ pp.\ \bibinfo {pages}
  {281--288}\BibitemShut {NoStop}%
%%CITATION = KFKI-75-45;%%
\bibitem [{\citenamefont {Ftaclas}\ and\ \citenamefont {Cohen}(1978)}]{FC78}%
  \BibitemOpen
  \bibfield  {author} {\bibinfo {author} {\bibfnamefont {C.}~\bibnamefont
  {Ftaclas}}\ and\ \bibinfo {author} {\bibfnamefont {J.~M.}\ \bibnamefont
  {Cohen}},\ }\href {\doibase 10.1103/PhysRevD.18.4373} {\bibfield  {journal}
  {\bibinfo  {journal} {Phys. Rev. D}\ }\textbf {\bibinfo {volume} {{\bf
  18}}},\ \bibinfo {pages} {4373} (\bibinfo {year} {1978})}\BibitemShut
  {NoStop}%
%%CITATION = PHRVA,D18,4373;%%
\bibitem [{\citenamefont {Born}(1934)}]{Born34}%
  \BibitemOpen
  \bibfield  {author} {\bibinfo {author} {\bibfnamefont {M.}~\bibnamefont
  {Born}},\ }\href {\doibase 10.1098/rspa.1934.0010} {\bibfield  {journal}
  {\bibinfo  {journal} {Proc. R. Soc.}\ }\textbf {\bibinfo {volume} {A {\bf
  143}}},\ \bibinfo {pages} {410} (\bibinfo {year} {1934})}\BibitemShut
  {NoStop}%
%%CITATION = PRSLA,A143,410;%%
\bibitem [{\citenamefont {Born}\ and\ \citenamefont {Infeld}(1934)}]{BI34}%
  \BibitemOpen
  \bibfield  {author} {\bibinfo {author} {\bibfnamefont {M.}~\bibnamefont
  {Born}}\ and\ \bibinfo {author} {\bibfnamefont {L.}~\bibnamefont {Infeld}},\
  }\href {\doibase 10.1098/rspa.1934.0059} {\bibfield  {journal} {\bibinfo
  {journal} {Proc. R. Soc.}\ }\textbf {\bibinfo {volume} {A {\bf 144}}},\
  \bibinfo {pages} {425} (\bibinfo {year} {1934})}\BibitemShut {NoStop}%
%%CITATION = PRSLA,A144,425;%%
\bibitem [{\citenamefont {Ay{\'o}n-Beato}\ and\ \citenamefont
  {Garc{\'i}a}(1998)}]{ABG98}%
  \BibitemOpen
  \bibfield  {author} {\bibinfo {author} {\bibfnamefont {E.}~\bibnamefont
  {Ay{\'o}n-Beato}}\ and\ \bibinfo {author} {\bibfnamefont {A.}~\bibnamefont
  {Garc{\'i}a}},\ }\href {\doibase 10.1103/PhysRevLett.80.5056} {\bibfield
  {journal} {\bibinfo  {journal} {Phys. Rev. Lett.}\ }\textbf {\bibinfo
  {volume} {{\bf 80}}},\ \bibinfo {pages} {5056} (\bibinfo {year} {1998})},\
  \Eprint {http://arxiv.org/abs/gr-qc/9911046} {arXiv:gr-qc/9911046 [gr-qc]}
  \BibitemShut {NoStop}%
%%CITATION = GR-QC/9911046;%%
\bibitem [{\citenamefont {Bronnikov}(2001)}]{Bronnikov00}%
  \BibitemOpen
  \bibfield  {author} {\bibinfo {author} {\bibfnamefont {K.~A.}\ \bibnamefont
  {Bronnikov}},\ }\href {\doibase 10.1103/PhysRevD.63.044005} {\bibfield
  {journal} {\bibinfo  {journal} {Phys. Rev. D}\ }\textbf {\bibinfo {volume}
  {{\bf 63}}},\ \bibinfo {pages} {044005} (\bibinfo {year} {2001})},\ \Eprint
  {http://arxiv.org/abs/gr-qc/0006014} {arXiv:gr-qc/0006014 [gr-qc]}
  \BibitemShut {NoStop}%
%%CITATION = GR-QC/0006014;%%
\bibitem [{\citenamefont {Burinskii}\ and\ \citenamefont
  {Hildebrandt}(2002)}]{BH02}%
  \BibitemOpen
  \bibfield  {author} {\bibinfo {author} {\bibfnamefont {A.}~\bibnamefont
  {Burinskii}}\ and\ \bibinfo {author} {\bibfnamefont {S.~R.}\ \bibnamefont
  {Hildebrandt}},\ }\href {\doibase 10.1103/PhysRevD.65.104017} {\bibfield
  {journal} {\bibinfo  {journal} {Phys. Rev. D}\ }\textbf {\bibinfo {volume}
  {{\bf 65}}},\ \bibinfo {pages} {104017} (\bibinfo {year} {2002})},\ \Eprint
  {http://arxiv.org/abs/hep-th/0202066} {arXiv:hep-th/0202066 [hep-th]}
  \BibitemShut {NoStop}%
%%CITATION = HEP-TH/0202066;%%
\bibitem [{\citenamefont {Garc{\'i}a-Salcedo}\ and\ \citenamefont
  {Bret{\'o}n}(2000)}]{GSB00}%
  \BibitemOpen
  \bibfield  {author} {\bibinfo {author} {\bibfnamefont {R.}~\bibnamefont
  {Garc{\'i}a-Salcedo}}\ and\ \bibinfo {author} {\bibfnamefont
  {N.}~\bibnamefont {Bret{\'o}n}},\ }\href {\doibase 10.1142/S0217751X00002160}
  {\bibfield  {journal} {\bibinfo  {journal} {Int. J. Mod. Phys.}\ }\textbf
  {\bibinfo {volume} {A {\bf 15}}},\ \bibinfo {pages} {4341} (\bibinfo {year}
  {2000})},\ \Eprint {http://arxiv.org/abs/gr-qc/0004017} {arXiv:gr-qc/0004017
  [gr-qc]} \BibitemShut {NoStop}%
%%CITATION = GR-QC/0004017;%%
\bibitem [{\citenamefont {Garc{\'i}a-Salcedo}\ and\ \citenamefont
  {Bret{\'o}n}(2005)}]{GSB04}%
  \BibitemOpen
  \bibfield  {author} {\bibinfo {author} {\bibfnamefont {R.}~\bibnamefont
  {Garc{\'i}a-Salcedo}}\ and\ \bibinfo {author} {\bibfnamefont
  {N.}~\bibnamefont {Bret{\'o}n}},\ }\href {\doibase
  10.1088/0264-9381/22/22/009} {\bibfield  {journal} {\bibinfo  {journal}
  {Class. Quantum Grav.}\ }\textbf {\bibinfo {volume} {{\bf 22}}},\ \bibinfo
  {pages} {4783} (\bibinfo {year} {2005})},\ \Eprint
  {http://arxiv.org/abs/gr-qc/0410142} {arXiv:gr-qc/0410142 [gr-qc]}
  \BibitemShut {NoStop}%
%%CITATION = GR-QC/0410142;%%
\bibitem [{\citenamefont {Camara}\ \emph {et~al.}(2004)\citenamefont {Camara},
  \citenamefont {de~Garcia~Maia}, \citenamefont {Carvalho},\ and\ \citenamefont
  {Lima}}]{CGMCL04}%
  \BibitemOpen
  \bibfield  {author} {\bibinfo {author} {\bibfnamefont {C.~S.}\ \bibnamefont
  {Camara}}, \bibinfo {author} {\bibfnamefont {M.~R.}\ \bibnamefont
  {de~Garcia~Maia}}, \bibinfo {author} {\bibfnamefont {J.~C.}\ \bibnamefont
  {Carvalho}}, \ and\ \bibinfo {author} {\bibfnamefont {J.~A.~S.}\ \bibnamefont
  {Lima}},\ }\href {\doibase 10.1103/PhysRevD.69.123504} {\bibfield  {journal}
  {\bibinfo  {journal} {Phys. Rev. D}\ }\textbf {\bibinfo {volume} {{\bf
  69}}},\ \bibinfo {pages} {123504} (\bibinfo {year} {2004})},\ \Eprint
  {http://arxiv.org/abs/astro-ph/0402311} {arXiv:astro-ph/0402311 [astro-ph]}
  \BibitemShut {NoStop}%
%%CITATION = ASTRO-PH/0402311;%%
\bibitem [{\citenamefont {Elizalde}\ \emph {et~al.}(2003)\citenamefont
  {Elizalde}, \citenamefont {Lidsey}, \citenamefont {Nojiri},\ and\
  \citenamefont {Odintsov}}]{ELNO03}%
  \BibitemOpen
  \bibfield  {author} {\bibinfo {author} {\bibfnamefont {E.}~\bibnamefont
  {Elizalde}}, \bibinfo {author} {\bibfnamefont {J.~E.}\ \bibnamefont
  {Lidsey}}, \bibinfo {author} {\bibfnamefont {S.}~\bibnamefont {Nojiri}}, \
  and\ \bibinfo {author} {\bibfnamefont {S.~D.}\ \bibnamefont {Odintsov}},\
  }\href {\doibase 10.1016/j.physletb.2003.08.074} {\bibfield  {journal}
  {\bibinfo  {journal} {Phys. Lett. B}\ }\textbf {\bibinfo {volume} {{\bf
  574}}},\ \bibinfo {pages} {1} (\bibinfo {year} {2003})},\ \Eprint
  {http://arxiv.org/abs/hep-th/0307177} {arXiv:hep-th/0307177 [hep-th]}
  \BibitemShut {NoStop}%
%%CITATION = HEP-TH/0307177;%%
\bibitem [{\citenamefont {Labun}\ and\ \citenamefont {Rafelski}(2010)}]{LR10}%
  \BibitemOpen
  \bibfield  {author} {\bibinfo {author} {\bibfnamefont {L.}~\bibnamefont
  {Labun}}\ and\ \bibinfo {author} {\bibfnamefont {J.}~\bibnamefont
  {Rafelski}},\ }\href {\doibase 10.1103/PhysRevD.81.065026} {\bibfield
  {journal} {\bibinfo  {journal} {Phys. Rev. D}\ }\textbf {\bibinfo {volume}
  {{\bf 81}}},\ \bibinfo {pages} {065026} (\bibinfo {year} {2010})},\ \Eprint
  {http://arxiv.org/abs/0811.4467} {arXiv:0811.4467 [hep-th]} \BibitemShut
  {NoStop}%
%%CITATION = ARXIV:0811.4467;%%
\bibitem [{\citenamefont {Heisenberg}\ and\ \citenamefont
  {Euler}(1936)}]{HE36}%
  \BibitemOpen
  \bibfield  {author} {\bibinfo {author} {\bibfnamefont {W.}~\bibnamefont
  {Heisenberg}}\ and\ \bibinfo {author} {\bibfnamefont {H.~Z.}\ \bibnamefont
  {Euler}},\ }\href {\doibase 10.1007/BF01343663} {\bibfield  {journal}
  {\bibinfo  {journal} {Z. Phys.}\ }\textbf {\bibinfo {volume} {{\bf 98}}},\
  \bibinfo {pages} {714} (\bibinfo {year} {1936})},\ \Eprint
  {http://arxiv.org/abs/physics/0605038} {arXiv:physics/0605038 [physics]}
  \BibitemShut {NoStop}%
%%CITATION = PHYSICS/0605038;%%
\bibitem [{\citenamefont {Fradkin}\ and\ \citenamefont
  {Tseytlin}(1985)}]{FT85}%
  \BibitemOpen
  \bibfield  {author} {\bibinfo {author} {\bibfnamefont {E.~S.}\ \bibnamefont
  {Fradkin}}\ and\ \bibinfo {author} {\bibfnamefont {A.~A.}\ \bibnamefont
  {Tseytlin}},\ }\href {\doibase 10.1016/0370-2693(85)90205-9} {\bibfield
  {journal} {\bibinfo  {journal} {Phys. Lett.}\ }\textbf {\bibinfo {volume}
  {{\bf 163B}}},\ \bibinfo {pages} {123} (\bibinfo {year} {1985})}\BibitemShut
  {NoStop}%
%%CITATION = PHLTA,B163,123;%%
\bibitem [{\citenamefont {Battesti}\ and\ \citenamefont {Rizzo}(2013)}]{BR13}%
  \BibitemOpen
  \bibfield  {author} {\bibinfo {author} {\bibfnamefont {R.}~\bibnamefont
  {Battesti}}\ and\ \bibinfo {author} {\bibfnamefont {C.}~\bibnamefont
  {Rizzo}},\ }\href {\doibase 10.1088/0034-4885/76/1/016401} {\bibfield
  {journal} {\bibinfo  {journal} {Rept. Prog. Phys.}\ }\textbf {\bibinfo
  {volume} {{\bf 76}}},\ \bibinfo {pages} {016401} (\bibinfo {year} {2013})},\
  \Eprint {http://arxiv.org/abs/1211.1933} {arXiv:1211.1933 [physics.optics]}
  \BibitemShut {NoStop}%
%%CITATION = ARXIV:1211.1933;%%
\bibitem [{\citenamefont {Fouch{\'e}}\ \emph {et~al.}(2016)\citenamefont
  {Fouch{\'e}}, \citenamefont {Battesti},\ and\ \citenamefont {Rizzo}}]{FBR16}%
  \BibitemOpen
  \bibfield  {author} {\bibinfo {author} {\bibfnamefont {M.}~\bibnamefont
  {Fouch{\'e}}}, \bibinfo {author} {\bibfnamefont {R.}~\bibnamefont
  {Battesti}}, \ and\ \bibinfo {author} {\bibfnamefont {C.}~\bibnamefont
  {Rizzo}},\ }\href {\doibase 10.1103/PhysRevD.93.093020} {\bibfield  {journal}
  {\bibinfo  {journal} {Phys. Rev. D}\ }\textbf {\bibinfo {volume} {{\bf
  93}}},\ \bibinfo {pages} {093020} (\bibinfo {year} {2016})}\BibitemShut
  {NoStop}%
\bibitem [{\citenamefont {Mosquera~Cuesta}\ \emph {et~al.}(2017)\citenamefont
  {Mosquera~Cuesta}, \citenamefont {Lambiase},\ and\ \citenamefont
  {Pereira}}]{MCLP17}%
  \BibitemOpen
  \bibfield  {author} {\bibinfo {author} {\bibfnamefont {H.~J.}\ \bibnamefont
  {Mosquera~Cuesta}}, \bibinfo {author} {\bibfnamefont {G.}~\bibnamefont
  {Lambiase}}, \ and\ \bibinfo {author} {\bibfnamefont {J.~P.}\ \bibnamefont
  {Pereira}},\ }\href {\doibase 10.1103/PhysRevD.95.025011} {\bibfield
  {journal} {\bibinfo  {journal} {Phys. Rev. D}\ }\textbf {\bibinfo {volume}
  {{\bf 95}}},\ \bibinfo {pages} {025011} (\bibinfo {year} {2017})},\ \Eprint
  {http://arxiv.org/abs/1701.00431} {arXiv:1701.00431 [gr-qc]} \BibitemShut
  {NoStop}%
%%CITATION = ARXIV:1701.00431;%%
\bibitem [{\citenamefont {Ellis}\ \emph {et~al.}(2017)\citenamefont {Ellis},
  \citenamefont {Mavromatos},\ and\ \citenamefont {You}}]{EMY17}%
  \BibitemOpen
  \bibfield  {author} {\bibinfo {author} {\bibfnamefont {J.}~\bibnamefont
  {Ellis}}, \bibinfo {author} {\bibfnamefont {N.~E.}\ \bibnamefont
  {Mavromatos}}, \ and\ \bibinfo {author} {\bibfnamefont {T.}~\bibnamefont
  {You}},\ }\href@noop {} {\enquote {\bibinfo {title} {{Light-by-Light
  Scattering Constraint on Born-Infeld Theory}},}\ } (\bibinfo {year} {2017}),\
  \Eprint {http://arxiv.org/abs/1703.08450} {arXiv:1703.08450 [hep-ph]}
  \BibitemShut {NoStop}%
%%CITATION = ARXIV:1703.08450;%%
\bibitem [{\citenamefont {Rasheed}(1997)}]{Rasheed97}%
  \BibitemOpen
  \bibfield  {author} {\bibinfo {author} {\bibfnamefont {D.~A.}\ \bibnamefont
  {Rasheed}},\ }\href@noop {} {\enquote {\bibinfo {title} {{Nonlinear
  electrodynamics: Zeroth and first laws of black hole mechanics}},}\ }
  (\bibinfo {year} {1997}),\ \Eprint {http://arxiv.org/abs/hep-th/9702087}
  {arXiv:hep-th/9702087 [hep-th]} \BibitemShut {NoStop}%
%%CITATION = HEP-TH/9702087;%%
\bibitem [{\citenamefont {Garcia}\ \emph {et~al.}(2012)\citenamefont {Garcia},
  \citenamefont {Hackmann}, \citenamefont {L{\"a}mmerzahl},\ and\ \citenamefont
  {Mac{\'i}as}}]{GHLM12}%
  \BibitemOpen
  \bibfield  {author} {\bibinfo {author} {\bibfnamefont {A.}~\bibnamefont
  {Garcia}}, \bibinfo {author} {\bibfnamefont {E.}~\bibnamefont {Hackmann}},
  \bibinfo {author} {\bibfnamefont {C.}~\bibnamefont {L{\"a}mmerzahl}}, \ and\
  \bibinfo {author} {\bibfnamefont {A.}~\bibnamefont {Mac{\'i}as}},\ }\href
  {\doibase 10.1103/PhysRevD.86.024037} {\bibfield  {journal} {\bibinfo
  {journal} {Phys. Rev. D}\ }\textbf {\bibinfo {volume} {{\bf 86}}},\ \bibinfo
  {pages} {024037} (\bibinfo {year} {2012})}\BibitemShut {NoStop}%
%%CITATION = PHRVA,D86,024037;%%
\bibitem [{\citenamefont {Cao}\ \emph {et~al.}(2014)\citenamefont {Cao},
  \citenamefont {Peng},\ and\ \citenamefont {Xu}}]{CPX14}%
  \BibitemOpen
  \bibfield  {author} {\bibinfo {author} {\bibfnamefont {L.-M.}\ \bibnamefont
  {Cao}}, \bibinfo {author} {\bibfnamefont {Y.}~\bibnamefont {Peng}}, \ and\
  \bibinfo {author} {\bibfnamefont {J.}~\bibnamefont {Xu}},\ }\href {\doibase
  10.1103/PhysRevD.90.024046} {\bibfield  {journal} {\bibinfo  {journal} {Phys.
  Rev. D}\ }\textbf {\bibinfo {volume} {{\bf 90}}},\ \bibinfo {pages} {024046}
  (\bibinfo {year} {2014})},\ \Eprint {http://arxiv.org/abs/1404.6639}
  {arXiv:1404.6639 [gr-qc]} \BibitemShut {NoStop}%
%%CITATION = ARXIV:1404.6639;%%
\bibitem [{\citenamefont {Herdeiro}\ and\ \citenamefont {Radu}(2015)}]{HR15}%
  \BibitemOpen
  \bibfield  {author} {\bibinfo {author} {\bibfnamefont {C.~A.~R.}\
  \bibnamefont {Herdeiro}}\ and\ \bibinfo {author} {\bibfnamefont
  {E.}~\bibnamefont {Radu}},\ }\bibfield  {booktitle} {\emph {\bibinfo
  {booktitle} {{Proceedings, 7th Black Holes Workshop 2014}}},\ }\href
  {\doibase 10.1142/S0218271815420146} {\bibfield  {journal} {\bibinfo
  {journal} {Int. J. Mod. Phys. D}\ }\textbf {\bibinfo {volume} {{\bf 24}}},\
  \bibinfo {pages} {1542014} (\bibinfo {year} {2015})},\ \Eprint
  {http://arxiv.org/abs/1504.08209} {arXiv:1504.08209 [gr-qc]} \BibitemShut
  {NoStop}%
%%CITATION = ARXIV:1504.08209;%%
\bibitem [{\citenamefont {Smoli{\'c}}(2017)}]{ISm17}%
  \BibitemOpen
  \bibfield  {author} {\bibinfo {author} {\bibfnamefont {I.}~\bibnamefont
  {Smoli{\'c}}},\ }\href {\doibase 10.1103/PhysRevD.95.024016} {\bibfield
  {journal} {\bibinfo  {journal} {Phys. Rev. D}\ }\textbf {\bibinfo {volume}
  {{\bf 95}}},\ \bibinfo {pages} {024016} (\bibinfo {year} {2017})},\ \Eprint
  {http://arxiv.org/abs/1609.04013} {arXiv:1609.04013 [gr-qc]} \BibitemShut
  {NoStop}%
%%CITATION = ARXIV:1609.04013;%%
\bibitem [{\citenamefont {Garc{\'i}a}\ \emph {et~al.}(1984)\citenamefont
  {Garc{\'i}a}, \citenamefont {Salazar},\ and\ \citenamefont
  {Pleba{\'n}ski}}]{GSP84}%
  \BibitemOpen
  \bibfield  {author} {\bibinfo {author} {\bibfnamefont {D.~A.}\ \bibnamefont
  {Garc{\'i}a}}, \bibinfo {author} {\bibfnamefont {I.~H.}\ \bibnamefont
  {Salazar}}, \ and\ \bibinfo {author} {\bibfnamefont {J.~F.}\ \bibnamefont
  {Pleba{\'n}ski}},\ }\href {\doibase 10.1007/BF02721649} {\bibfield  {journal}
  {\bibinfo  {journal} {Nuovo Cimento B Serie}\ }\textbf {\bibinfo {volume}
  {{\bf 84}}},\ \bibinfo {pages} {65} (\bibinfo {year} {1984})}\BibitemShut
  {NoStop}%
\bibitem [{\citenamefont {Hassa{\"i}ne}\ and\ \citenamefont
  {Mart{\'i}nez}(2007)}]{HM07}%
  \BibitemOpen
  \bibfield  {author} {\bibinfo {author} {\bibfnamefont {M.}~\bibnamefont
  {Hassa{\"i}ne}}\ and\ \bibinfo {author} {\bibfnamefont {C.}~\bibnamefont
  {Mart{\'i}nez}},\ }\href {\doibase 10.1103/PhysRevD.75.027502} {\bibfield
  {journal} {\bibinfo  {journal} {Phys. Rev. D}\ }\textbf {\bibinfo {volume}
  {{\bf 75}}},\ \bibinfo {pages} {027502} (\bibinfo {year} {2007})},\ \Eprint
  {http://arxiv.org/abs/hep-th/0701058} {arXiv:hep-th/0701058 [hep-th]}
  \BibitemShut {NoStop}%
%%CITATION = HEP-TH/0701058;%%
\bibitem [{\citenamefont {Hassa{\"i}ne}\ and\ \citenamefont
  {Mart{\'i}nez}(2008)}]{HM08}%
  \BibitemOpen
  \bibfield  {author} {\bibinfo {author} {\bibfnamefont {M.}~\bibnamefont
  {Hassa{\"i}ne}}\ and\ \bibinfo {author} {\bibfnamefont {C.}~\bibnamefont
  {Mart{\'i}nez}},\ }\href {\doibase 10.1088/0264-9381/25/19/195023} {\bibfield
   {journal} {\bibinfo  {journal} {Class. Quantum Grav.}\ }\textbf {\bibinfo
  {volume} {{\bf 25}}},\ \bibinfo {pages} {195023} (\bibinfo {year} {2008})},\
  \Eprint {http://arxiv.org/abs/0803.2946} {arXiv:0803.2946 [hep-th]}
  \BibitemShut {NoStop}%
%%CITATION = ARXIV:0803.2946;%%
\bibitem [{\citenamefont {Yajima}\ and\ \citenamefont {Tamaki}(2001)}]{YT00}%
  \BibitemOpen
  \bibfield  {author} {\bibinfo {author} {\bibfnamefont {H.}~\bibnamefont
  {Yajima}}\ and\ \bibinfo {author} {\bibfnamefont {T.}~\bibnamefont
  {Tamaki}},\ }\href {\doibase 10.1103/PhysRevD.63.064007} {\bibfield
  {journal} {\bibinfo  {journal} {Phys. Rev. D}\ }\textbf {\bibinfo {volume}
  {{\bf 63}}},\ \bibinfo {pages} {064007} (\bibinfo {year} {2001})},\ \Eprint
  {http://arxiv.org/abs/gr-qc/0005016} {arXiv:gr-qc/0005016 [gr-qc]}
  \BibitemShut {NoStop}%
%%CITATION = GR-QC/0005016;%%
\bibitem [{\citenamefont {Ruffini}\ \emph {et~al.}(2013)\citenamefont
  {Ruffini}, \citenamefont {Wu},\ and\ \citenamefont {Xue}}]{RWX13}%
  \BibitemOpen
  \bibfield  {author} {\bibinfo {author} {\bibfnamefont {R.}~\bibnamefont
  {Ruffini}}, \bibinfo {author} {\bibfnamefont {Y.-B.}\ \bibnamefont {Wu}}, \
  and\ \bibinfo {author} {\bibfnamefont {S.-S.}\ \bibnamefont {Xue}},\ }\href
  {\doibase 10.1103/PhysRevD.88.085004} {\bibfield  {journal} {\bibinfo
  {journal} {Phys. Rev. D}\ }\textbf {\bibinfo {volume} {{\bf 88}}},\ \bibinfo
  {pages} {085004} (\bibinfo {year} {2013})},\ \Eprint
  {http://arxiv.org/abs/1307.4951} {arXiv:1307.4951 [hep-th]} \BibitemShut
  {NoStop}%
%%CITATION = ARXIV:1307.4951;%%
\bibitem [{\citenamefont {Hoffmann}\ and\ \citenamefont {Infeld}(1937)}]{HI37}%
  \BibitemOpen
  \bibfield  {author} {\bibinfo {author} {\bibfnamefont {B.}~\bibnamefont
  {Hoffmann}}\ and\ \bibinfo {author} {\bibfnamefont {L.}~\bibnamefont
  {Infeld}},\ }\href {\doibase 10.1103/PhysRev.51.765} {\bibfield  {journal}
  {\bibinfo  {journal} {Phys. Rev.}\ }\textbf {\bibinfo {volume} {{\bf 51}}},\
  \bibinfo {pages} {765} (\bibinfo {year} {1937})}\BibitemShut {NoStop}%
%%CITATION = PHRVA,51,765;%%
\bibitem [{\citenamefont {Aiello}\ \emph {et~al.}(2005)\citenamefont {Aiello},
  \citenamefont {Ferraro},\ and\ \citenamefont {Giribet}}]{AFG05}%
  \BibitemOpen
  \bibfield  {author} {\bibinfo {author} {\bibfnamefont {M.}~\bibnamefont
  {Aiello}}, \bibinfo {author} {\bibfnamefont {R.}~\bibnamefont {Ferraro}}, \
  and\ \bibinfo {author} {\bibfnamefont {G.}~\bibnamefont {Giribet}},\ }\href
  {\doibase 10.1088/0264-9381/22/13/004} {\bibfield  {journal} {\bibinfo
  {journal} {Class. Quantum Grav.}\ }\textbf {\bibinfo {volume} {22}},\
  \bibinfo {pages} {2579} (\bibinfo {year} {2005})},\ \Eprint
  {http://arxiv.org/abs/gr-qc/0502069} {arXiv:gr-qc/0502069 [gr-qc]}
  \BibitemShut {NoStop}%
%%CITATION = GR-QC/0502069;%%
\bibitem [{\citenamefont {Soleng}(1995)}]{Soleng95}%
  \BibitemOpen
  \bibfield  {author} {\bibinfo {author} {\bibfnamefont {H.~H.}\ \bibnamefont
  {Soleng}},\ }\href {\doibase 10.1103/PhysRevD.52.6178} {\bibfield  {journal}
  {\bibinfo  {journal} {Phys. Rev. D}\ }\textbf {\bibinfo {volume} {{\bf
  52}}},\ \bibinfo {pages} {6178} (\bibinfo {year} {1995})},\ \Eprint
  {http://arxiv.org/abs/hep-th/9509033} {arXiv:hep-th/9509033 [hep-th]}
  \BibitemShut {NoStop}%
%%CITATION = HEP-TH/9509033;%%
\bibitem [{\citenamefont {Hendi}(2012)}]{Hendi12}%
  \BibitemOpen
  \bibfield  {author} {\bibinfo {author} {\bibfnamefont {S.~H.}\ \bibnamefont
  {Hendi}},\ }\href {\doibase 10.1007/JHEP03(2012)065} {\bibfield  {journal}
  {\bibinfo  {journal} {JHEP}\ }\textbf {\bibinfo {volume} {{\bf 03}}},\
  \bibinfo {pages} {065} (\bibinfo {year} {2012})},\ \Eprint
  {http://arxiv.org/abs/1405.4941} {arXiv:1405.4941 [hep-th]} \BibitemShut
  {NoStop}%
%%CITATION = ARXIV:1405.4941;%%
\bibitem [{\citenamefont {Kruglov}(2015)}]{Kruglov15}%
  \BibitemOpen
  \bibfield  {author} {\bibinfo {author} {\bibfnamefont {S.~I.}\ \bibnamefont
  {Kruglov}},\ }\href {\doibase 10.1016/j.aop.2014.12.001} {\bibfield
  {journal} {\bibinfo  {journal} {Annals of Physics}\ }\textbf {\bibinfo
  {volume} {{\bf 353}}},\ \bibinfo {pages} {299} (\bibinfo {year} {2015})},\
  \Eprint {http://arxiv.org/abs/1410.0351} {arXiv:1410.0351 [physics.gen-ph]}
  \BibitemShut {NoStop}%
\bibitem [{\citenamefont {Gibbons}\ and\ \citenamefont
  {Herdeiro}(2001)}]{GH01}%
  \BibitemOpen
  \bibfield  {author} {\bibinfo {author} {\bibfnamefont {G.~W.}\ \bibnamefont
  {Gibbons}}\ and\ \bibinfo {author} {\bibfnamefont {C.~A.~R.}\ \bibnamefont
  {Herdeiro}},\ }\href {\doibase 10.1103/PhysRevD.63.064006} {\bibfield
  {journal} {\bibinfo  {journal} {Phys. Rev. D}\ }\textbf {\bibinfo {volume}
  {{\bf 63}}},\ \bibinfo {pages} {064006} (\bibinfo {year} {2001})},\ \Eprint
  {http://arxiv.org/abs/hep-th/0008052} {arXiv:hep-th/0008052 [hep-th]}
  \BibitemShut {NoStop}%
%%CITATION = HEP-TH/0008052;%%
\bibitem [{\citenamefont {Banerjee}(1970)}]{Banerjee70}%
  \BibitemOpen
  \bibfield  {author} {\bibinfo {author} {\bibfnamefont {A.}~\bibnamefont
  {Banerjee}},\ }\href {\doibase 10.1063/1.1665071} {\bibfield  {journal}
  {\bibinfo  {journal} {J. Math. Phys.}\ }\textbf {\bibinfo {volume} {{\bf
  11}}},\ \bibinfo {pages} {51} (\bibinfo {year} {1970})}\BibitemShut {NoStop}%
\bibitem [{\citenamefont {Cvitan}\ \emph {et~al.}(2016)\citenamefont {Cvitan},
  \citenamefont {Dominis~Prester},\ and\ \citenamefont {Smoli{\'c}}}]{CDPS16}%
  \BibitemOpen
  \bibfield  {author} {\bibinfo {author} {\bibfnamefont {M.}~\bibnamefont
  {Cvitan}}, \bibinfo {author} {\bibfnamefont {P.}~\bibnamefont
  {Dominis~Prester}}, \ and\ \bibinfo {author} {\bibfnamefont {I.}~\bibnamefont
  {Smoli{\'c}}},\ }\href {\doibase 10.1088/0264-9381/33/7/077001} {\bibfield
  {journal} {\bibinfo  {journal} {Class. Quantum Grav.}\ }\textbf {\bibinfo
  {volume} {{\bf 33}}},\ \bibinfo {pages} {077001} (\bibinfo {year} {2016})},\
  \bibinfo {note}
  {\href{https://cqgplus.com/2016/03/23/perfect-accordance-of-the-gravitational-and-the-electromagnetic-field-in-3d/}{CQG+}},\
  \Eprint {http://arxiv.org/abs/1508.03343} {arXiv:1508.03343 [gr-qc]}
  \BibitemShut {NoStop}%
%%CITATION = ARXIV:1508.03343;%%
\bibitem [{\citenamefont {Krongos}\ and\ \citenamefont {Torre}(2017)}]{KT17}%
  \BibitemOpen
  \bibfield  {author} {\bibinfo {author} {\bibfnamefont {D.~S.}\ \bibnamefont
  {Krongos}}\ and\ \bibinfo {author} {\bibfnamefont {C.~G.}\ \bibnamefont
  {Torre}},\ }\href {\doibase 10.1063/1.4974091} {\bibfield  {journal}
  {\bibinfo  {journal} {J. Math. Phys.}\ }\textbf {\bibinfo {volume} {{\bf
  58}}},\ \bibinfo {pages} {012501} (\bibinfo {year} {2017})},\ \Eprint
  {http://arxiv.org/abs/1611.04143} {arXiv:1611.04143 [gr-qc]} \BibitemShut
  {NoStop}%
%%CITATION = ARXIV:1611.04143;%%
\bibitem [{\citenamefont {Smoli{\'c}}(2015)}]{ISm15}%
  \BibitemOpen
  \bibfield  {author} {\bibinfo {author} {\bibfnamefont {I.}~\bibnamefont
  {Smoli{\'c}}},\ }\href {\doibase 10.1088/0264-9381/32/14/145010} {\bibfield
  {journal} {\bibinfo  {journal} {Class. Quantum Grav.}\ }\textbf {\bibinfo
  {volume} {{\bf 32}}},\ \bibinfo {pages} {145010} (\bibinfo {year} {2015})},\
  \Eprint {http://arxiv.org/abs/1501.04967} {arXiv:1501.04967 [gr-qc]}
  \BibitemShut {NoStop}%
%%CITATION = ARXIV:1501.04967;%%
\bibitem [{\citenamefont {Smoli{\'c}}(2012)}]{ISm12}%
  \BibitemOpen
  \bibfield  {author} {\bibinfo {author} {\bibfnamefont {I.}~\bibnamefont
  {Smoli{\'c}}},\ }\href {\doibase 10.1088/0264-9381/29/20/207002} {\bibfield
  {journal} {\bibinfo  {journal} {Class. Quantum Grav.}\ }\textbf {\bibinfo
  {volume} {{\bf 29}}},\ \bibinfo {pages} {207002} (\bibinfo {year} {2012})},\
  \Eprint {http://arxiv.org/abs/1205.1071} {arXiv:1205.1071 [gr-qc]}
  \BibitemShut {NoStop}%
%%CITATION = ARXIV:1205.1071;%%
\bibitem [{\citenamefont {Smoli{\'c}}(2014)}]{ISm14}%
  \BibitemOpen
  \bibfield  {author} {\bibinfo {author} {\bibfnamefont {I.}~\bibnamefont
  {Smoli{\'c}}},\ }\href {\doibase 10.1088/0264-9381/31/23/235002} {\bibfield
  {journal} {\bibinfo  {journal} {Class. Quantum Grav.}\ }\textbf {\bibinfo
  {volume} {{\bf 31}}},\ \bibinfo {pages} {235002} (\bibinfo {year} {2014})},\
  \Eprint {http://arxiv.org/abs/1404.1936} {arXiv:1404.1936 [gr-qc]}
  \BibitemShut {NoStop}%
%%CITATION = ARXIV:1404.1936;%%
\end{thebibliography}%
%%%%%%%%%%%%%%%%%%%%%%%%%%%%
%%%%%%%%%%%%%%%%%%%%%%%%%%%%

\end{document}